# A Quick Introduction to the strong coupling regime of Cavity Quantum Electrodynamics: applications and fundamental quantum theory

Nathan D. Poulin[*]

Dec. 1, 2014

**Abstract**

Since the photon box gedanken experiments of several of the founding fathers of modern physics, considerable progress has been made in differentiating the quantum and classical worlds. In this pursuit, the cavity as an open quantum system has been indispensable. From the quantization of the atom and field within a superconducting cavity, a unique realm of EPR type entanglement has emerged. In this way, dynamical evolution of the system in the strong coupling regime is intimately tied with the coupling of an atom with a single resonant or non-resonant mode within the cavity. More specifically, the cavity can be prepared so that the atom is detected in a desired state. Here, the essentials of the strong coupling regime of Cavity Quantum Electrodynamics (QED) are reviewed for cavities tuned with a single atomic transition. A brief introduction of the systems is followed by an approach of the more striking effects, leading towards Ramsey Interferometry and Quantum Non-Demolition measurements as means for quantum gate protocol. Because the integrity of the atom and photon states is important for the advancement of quantum computation, a brief discussion of the decoherence problems is also presented. This document is meant to introduce the topic in a way that makes it easily accessible to those working in closely related areas of physics, and to highlight key applications and some basic questions concerning decoherence and the measurement problem.

## I. Introduction

Single photon and atom interactions are defining features of cavity quantum electrodynamics. When atoms are confined to superconducting cavities, their behavior is heavily influenced. Laser excitation of a group of atoms in free space can result in what are called Rabi oscillations. If we make the two level approximation, and the driving field is resonant with a single transition, then the ensemble of atoms has time varying (sinusoidal) probabilities of being in the excited (e) or ground (g) states. If the laser is pulsed, then the electric field varies in time. The pulses are described by their area, which correspond to an energy value. Each pulse can be thought of as an operation on the atom's state, visualized as a vector rotation around the Bloch sphere. An atom sent through a superconducting cavity with a single mode tuned to an atomic transition allows this type of interaction to occur with a single atom and photon. Other processes, like spontaneous emission rates, can be modified with great precision. But perhaps the most thought provoking aspects of the superconducting cavity are in its ability to harbor entangled states. Different pulses can suspend the passing atoms in any arbitrary superposition of their ground or excited states. These pulse operations, in conjunction with the coupling of the atom with the cavity field, can afford any number of complex entangled states. The realization of a controlled-NOT gate, as well as many other interesting experiments with the atom-cavity system, have many interested in furthering quantum information science. The advances of the field have also created a viable starting point for the development of new electro-optic devices.

---

[*] Previously affiliated with Union College Department of Physics and Astronomy (Schenectady, NY): email address: *poulinn91@gmail.com*

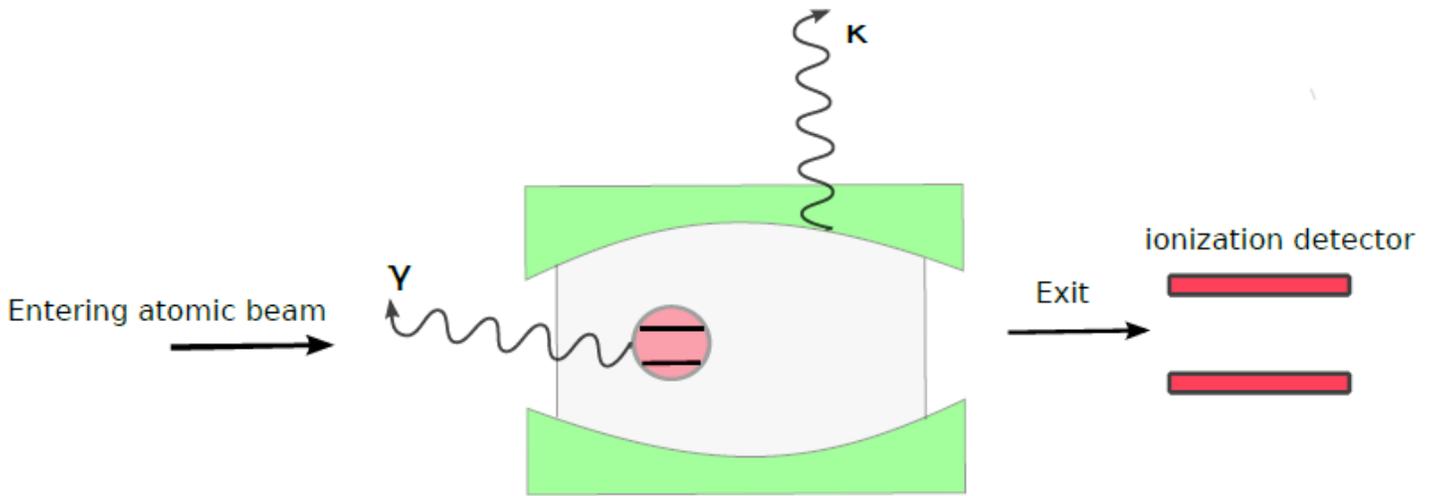

**Figure 1.** Schematic of a two-level atom within a resonant cavity, showing the photon decay (κ) and non-resonant decay (γ) parameters and the ionization detector. As the atom crosses the cavity, it interacts with the resonant mode (shaded).

The cavity is composed of two superconducting, spherical mirrors facing each other placed millimeters apart, seen in fig.1. A beam of atoms is sent across the cavity with an adjusted velocity and photons can be injected with a waveguide. The lifetime of photons within the cavity is an important parameter, and for years progress was limited by it. While the reflectivity of the mirrors is close to unity, photon leakage occurs on a time scale that limits the length of experiments. This decay process is associated with the photon decay rate (κ). The photon decay rate is determined from the quality factor ($Q$), where $\kappa = \omega/Q$, and ω is the resonant mode frequency. The dimensionless quality factor describes the energy loss from the cavity. The non-resonant decay rate (γ) is caused by spontaneous emission beyond the cavity mirror boundaries. If the atom crosses the cavity and spontaneously emits a photon, it is unlikely to reabsorb it if the photon and non-resonant decay rates are high—this is referred to as weak coupling. Many studies have been carried out in the weak regime, such as measurements of Lamb shifts and spontaneous emission effects. For the system to produce effects relevant to quantum computation and cryptography, the atom-cavity coupling constant $g_0$ must be much greater than κ and γ (Fox, 2006). This creates a system of reversible energy exchange.

## II. A Brief History

During the early days of Cavity QED (early 80s), much effort went into machining mirrors capable of obtaining the strong coupling limit, and in the words of Serge Haroche, this limit became their "Holy Grail." Initial breakthroughs were made by Meschede, Walther, and Muller at the Max-Planck institute, where they observed the exchange of photons between a cavity mode tuned to resonance with a single transition and passing atoms (Haroche, 2013). They engineered a closed cylinder structure, with the cavity mirrors housed within it. Since there were only small apertures for atom entry and exit, the non-resonant decay rate was significantly reduced. By sending atoms across the structure, they realized photons building up within the cavity. This can be understood as a "one-atom maser," (Meschede, Walther, and Müller, 1985) whereby the atom traverses the cavity with a particular velocity and has probabilities of exiting unchanged or having spontaneously emitted a photon. Eventually an atom will donate a photon to the cavity, and the next atom in the train will have an increased probability of doing so as well. This is because the rate at which the atom and field modes exchange energy is proportional to the number of photons within the cavity (refer to the Purcell effect) (Fox, 2006). The photons will continue to build up in the cavity until equilibrium is reached. The atoms that emerge will be in one of two possible states (Haroche and Raimond, 1993).

However, the closed structure was the cause for further technical difficulties. Atoms had to pass through tiny holes, very close to metal surfaces. Stray electric fields were produced by these surfaces, which perturb the atoms' states enough to destroy any prepared coherence. The École Normale Supérieure (ENS) group, with the help from scientists at the Center of Atomic Energy (CEA), was able to remedy the situation without abandoning their open cavity structure. By machining mirrors out of copper and doping a small amount of superconducting niobium on top of them, photon lifetimes reached 130 ms, travelling 40,000 km within the cavity (Haroche, 2013). The strong coupling regime had been reached, opening the door for a range of experiments with atoms of similar lifetimes—Rydberg atoms.

## III. Rydberg Atoms

Conventional two level atoms of the ground and first excited states have radiative lifetimes on the order of nanoseconds, being far too short for any practical use. Many experiments have been made possible by implementing highly excited atoms, formally known as Rydberg atoms. Their long radiative lifetimes are needed to extend the atom-cavity coupling effects.

To illustrate these atoms, let us recall the Bohr model. The principal quantum number *n* determines the atom's energy level, with the electrons orbiting the nucleus. Now consider a highly excited atom other than hydrogen. The outermost electron will take up an orbit that is well beyond that of the other electrons. From its perspective, the nucleus and inner electrons will appear as a core carrying a charge of +1, creating the Coulomb force needed to pull it into orbit. Since Rydberg atoms behave similarly to hydrogen, the electron is prevented from colliding with the nucleus—it can only occupy certain stationary states. Their single electrons are excited to *n* values of 50 or more, giving them atomic radii 1000's of times larger than a ground state atom. The rotating dipole acts as a huge antenna, allowing ease of interaction with the cavity mode.

With transitional energies varying as $1/n^3$, Rydberg atoms have relatively continuous emission energies for high *n*. In these highly excited states, spontaneous transitions to neighboring Rydberg levels are slower. If thought classically, emitted microwave photons are accompanied by the electron's successive hops from higher energy levels to slightly lower ones. Pulsed tunable dye lasers can discriminate between neighboring Rydberg states, leaving the choice for excitation to particular *n*, *l*, and *m* sub-levels. Typically, the Rydberg states are prepared in the highest angular momentum state, and thus are called circular Rydberg atoms. Preparing these atoms requires a complicated process, and the details have not been included here. In the scheme used by Kleppner et. al. (1981), they use two different pulses to excite the atom into an intermediate state, and a third pulse to send it into the Rydberg state. This has proven to be a highly efficient method for Rydberg preparation.

These fragile atoms can break apart from interatomic collisions. This is minimized by preparing them in an atomic beam, allowing for adequate spacing. The alkali metals used for the experiments are heated in an oven, vaporized, and sent through a small pore into a vacuum chamber. The vacuum prevents collisions, and after passing through another small aperture, the emerging beam is properly collimated. Although these atoms are electrically neutral, they are detected by ionization. Two field plates are laid above and below the beam and a pulse of voltage is stimulated between them. Relatively weak electric fields are sufficient enough to ionize the Rydberg atoms. The freed electron is then incident on a detector (Kleppner, Littman, and Zimmerman, 1981; Gallagher, 2005).

## IV. Mode Splitting and Vacuum Rabi Oscillations

If we prepare an arbitrary two level atom (in reality a Rydberg state) in an excited state e, and send it through an empty cavity, an effect analogous to the Zeeman shift for spin-½ systems can be observed. The cavity mode (ω) is tuned to resonance with the e to g transition ($\omega_{eg}$). The cavity contains the vacuum expressed in Dirac notation by |0⟩, and thus contains the zero-point energy of (1/2)ℏω. For ease of argument, we will consider the first excited state so that the combined atom and photon energy is (3/2)ℏω. As the atom traverses the cavity, it will have a probability of emitting and reabsorbing a reflected photon. We see that the atom-cavity system oscillates between two modes—an e atom with no photon and a g atom with 1 photon. These states can be expressed by |$e, 0$⟩ and |$g, 1$⟩ respectively, with the first entry of each ket representing the state and the second entry the number of

photons in the cavity. One would typically consider these as degenerate states, but the dressed atom formalism which follows from the Jaynes-Cummings model tells us that we can no longer think of the atom and photon as separate entities. This atom-cavity coupling removes the degeneracy, creating a doublet of energies of $(3/2)\hbar\omega \pm \hbar g_0$. While the preceding has described the first excited state, the argument can be extended for any nth excited level. Each level contains an energy splitting of $\Delta E_n = 2\sqrt{n}\hbar g_0$. What results is a series of doublets, with each doublet being a rung of the Jaynes-Cummings ladder. The difference between the doublet peaks is the oscillating frequency between modes.

Thompson, Rempe, and Kimble (1992) were among the first to observe this splitting for a single atom in a high-$Q$ optical cavity, by measuring the transmission intensity of the cavity. Although we have described microwave transitions with Rydberg atoms as the champions of strong coupling, certain dipole and quadrupole transitions in the optical regime have been used to achieve this limit; this domain has been pioneered by the Kimble group out of Caltech (1998). As you can see from fig.2, there is a single resonance peak for the vacuum mode. The coupling of a single atom with the cavity mode lifts the degeneracy.

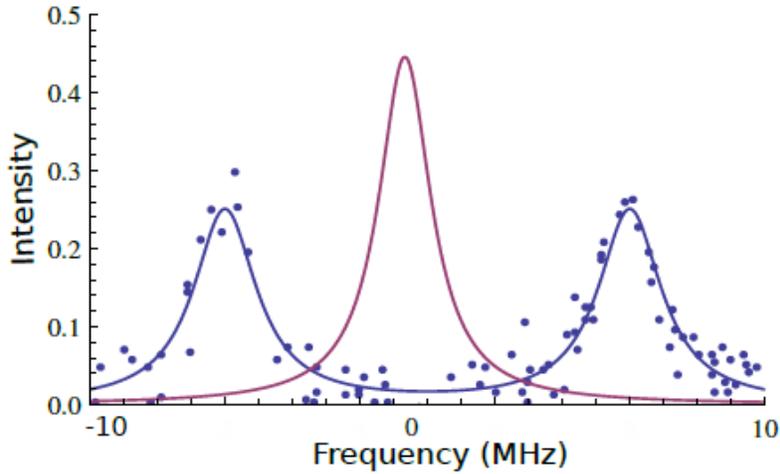

**Figure 2**. Transmission intensity with respect to rabi frequency for an empty cavity (purple), showing its single resonance peak from the vacuum energy, with depiction of mode splitting caused by a single atom within the cavity on average (blue). The intensity units are arbitrary. The plot was computer generated. See (Kimble, 1994) for experimental data.

The story is not over for this simple system—more information can be obtained by varying the duration of the atom-cavity coupling. For this next example and all of those that follow, we will be considering g and e Rydberg states for the n=50 and 51 levels. An atom at the center of the cavity with v=0, in $|e, 0\rangle$, evolves according to:

$$|\Psi_e(t)\rangle = \cos\left(\frac{\Omega t}{2}\right)|e, 0\rangle + \sin\left(\frac{\Omega t}{2}\right)|g, 1\rangle \quad (1)$$

With $\Omega/2$ being the atom-field coupling constant. This is a modified version of shining laser light on a group of atoms. In this case, the interaction involves single atoms and photons. If instead the initial state is $|g, 1\rangle$, time evolution of the system follows from:

$$|\Psi_g(t)\rangle = \cos\left(\frac{\Omega t}{2}\right)|g, 1\rangle - \sin\left(\frac{\Omega t}{2}\right)|e, 0\rangle \quad (2)$$

To better describe these oscillations, we need to introduce the spatial variability of the coupling interaction. This is accounted for by the interaction time $t_i$, given by:

$$t_i = \sqrt{\pi}w/v \quad (3)$$

with v being the velocity of the atom across the cavity. This $t_i$ represents a resonant interaction across the full mode and simply replaces t in the above wave functions. The atom that leaves the cavity mode, depending on the length of the interaction, can be suspended in a superposition of states. If the atom is detected, the mode in the cavity instantaneously collapses into its corresponding state, per the Copenhagen interpretation.

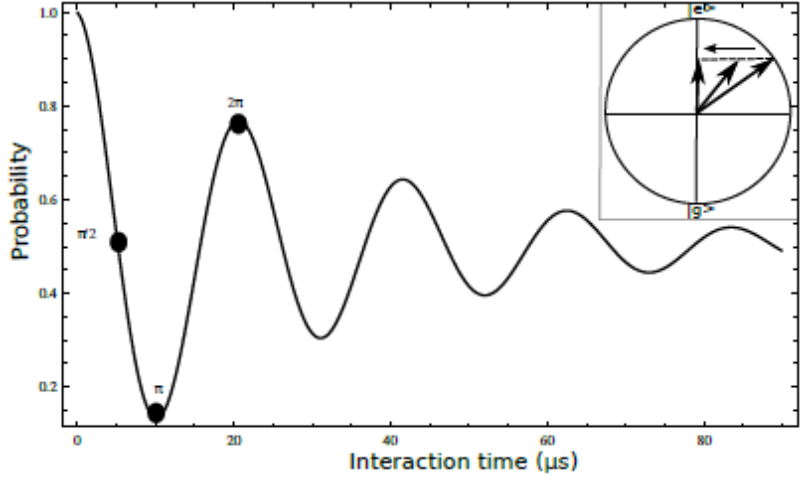

**Figure 3.** Vacuum rabi oscillations with the atom entering an empty resonant cavity in state e. The plot is a function of the probability $P_e$ of finding the atom in e with respect to the interaction time $t_i$. Three crucial times corresponding to the π/2, π, and 2π rabi pulses are shown. Damping is primarily due to transverse relaxation, leaving the relative populations intact, depicted in the inset of the bloch sphere. The north and south poles of the sphere represent the e and g pure states respectively, with superposition states in between. The azimuthal coordinate relays phase information of the dipole (not shown, see sec VI.). The plot is generated to depict the data from (Brune et al., 1996).

Brune et al. (1996) observed these Rabi Oscillations by measuring the probability of finding the atom in level e at time $t_i$, given by:

$$P_e = \frac{1+cos\Omega t_i}{2} \qquad (4)$$

With $t_i$ determined by the atomic velocity. As you can see from fig. 3, the probability oscillates at $\Omega/2\pi$. Many measurements were made over a range of interaction times (changed by varying atomic velocity), and then averaged. The data closely fit the curve in fig. 3. The resonant mode frequency ($\Omega$) multiplied by the interaction time ($t_i$) is a product that is used to describe the operation on the atom's state. Three important rotations are shown on fig. 3 (π/2, π, and 2π). These will become clearer in the next section. Because the interaction times are on the order of μs and the lifetimes of the atoms in the ms range, it is reasonable to assume that damping due to population decay is negligible. Transverse relaxation is ultimately what dampens the signal, caused by dark counts in the ionization detectors from thermal photons (Brune et al., 1996). This means that the superposition ends up behaving like a statistical mixture at increasing times. Thus, the Bloch vector shortens in length but its projection onto the vertical axis is maintained, shown on the inset of figure 3.

### V. Essential Rabi Pulses

Certain Rabi pulses are of particular interest. An equally weighted superposition state can be produced by what is called a "π/2 pulse." This occurs when $\Omega t_i = \pi/2$, and the state is:

$$|\Psi_{\pi/2}\rangle = \frac{1}{\sqrt{2}}(|e,0\rangle + |g,1\rangle) \qquad (5)$$

This is the simplest EPR entanglement scenario. Detection of the atom in e signals the cavity in vacuum, without ambiguity. Likewise, a detection of the atom in g signals the presence of a single photon in the cavity. A π-pulse applied to the $|e,0\rangle$ state will end up in the nonentangled $|g,1\rangle$ state. More interestingly, if the atom is already in an e and g superposition state upon entering, the atom leaves in the g state, with the cavity containing a superposition of the zero and one photon Fock states:

$$(c_e|e\rangle + c_g|g\rangle)|0\rangle \rightarrow |g\rangle(c_e|1\rangle + c_g|0\rangle) \qquad (6)$$

With $c_e$ and $c_g$ being the coefficients for each state. The square of the absolute value of a particular coefficient gives the configurations' probability of being in that state. The reverse, for an atom in the g state interacting with the cavity field in a coherent superposition of the 0 and 1 photon Fock state is given by:

$$(c_1|1\rangle + c_0|0\rangle)|g\rangle \to |0\rangle(-c_1|e\rangle + c_0|g\rangle) \quad (7)$$

A $2\pi$-pulse on either initial states e or g only introduces a global quantum phase shift of $\pi$:

$$|e,0\rangle \to -|e,0\rangle, \quad |g,1\rangle \to -|g,1\rangle \quad (8)$$

However a g state atom entering a cavity in vacuum experiences no such phase shift. The different pulses mentioned are important for the engineering of quantum logic gates, and will be applied in later sections (Raimond, Brune, Haroche, 2001; Rauschenbeutel, 2000).

## VI. Ramsey Interferometer

The Ramsey interferometer is very useful for introducing any number of entangled states. It applies pulses before and after the cavity interaction. The radiation is sent through the regions $R_1$ and $R_2$ (refer to fig. 4), produced by $S$ and funneled on both ends of the cavity mode. These fields have very short radiative lifetimes, in effect acting classically, and can be neglected in the discussion of entanglement between the cavity field and atomic qubits. They are essentially hadamard transformations. A more detailed review of Ramsey interferometry can be seen in (Orzel, 2012).

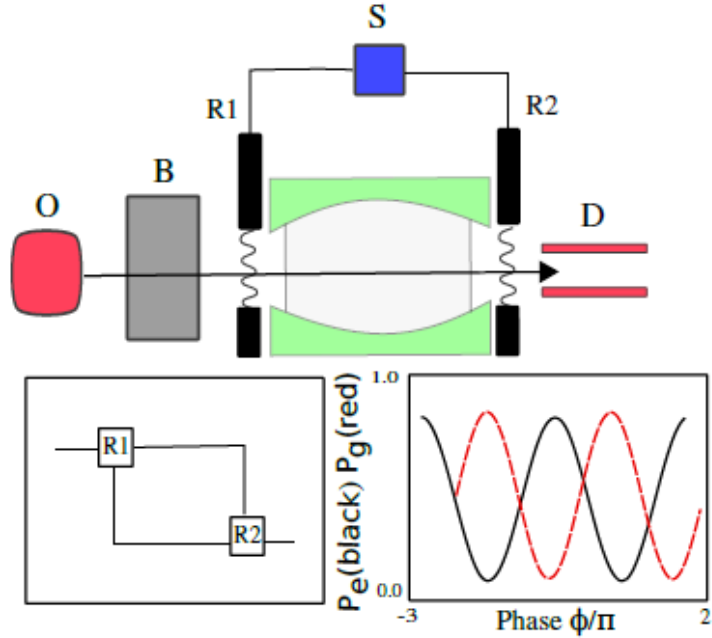

**Figure 4.** Schematic of experimental apparatus. Oven O velocity selects rubidium atoms, where they are transformed into Rydberg atoms in zone B. They travel across the superconducting cavity and are detected by field ionization at D. S can be used to inject a coherent field, which feeds an auxiliary source to R1 and R2. The schematic in the bottom left shows the different "paths" the atoms take through the interferometer, similar to a Mach-Zender Interferometer. The inset on the right is a plot of the probability of detecting the atom in g or e over a range of relative phases. See (Raimond, Brune, and Haroche, 2001).

If we again consider a two level atom, and send $\pi/2$ pulses through $R_1$ and $R_2$ at a frequency $\omega$ nearly resonant with a transition frequency $\omega_{eg}$, we can observe phase contrast (at this point we are ignoring the cavity mode). In other words, the Ramsey pulses act as beamsplitters of a Mach-Zender interferometer, seen in fig. 4. The atom must have a dipole to strongly interact with the cavity modes. The excited state Rydberg atoms containing a symmetric charge distribution before being hit by $R_1$, are characterized by a de Broglie wavelength of uniform amplitude with an integer number of wavelengths around the circumference. The $R_1$ microwave pulse can be applied, so that the atoms are in a superposition of g and e Rydberg states. The sum of the two de Broglie wavelengths of this superposition state has a node at one point of its orbit, resulting from destructive interference. Likewise, the other end of the orbit contains a maximum charge density by constructive interference, seen in fig. 5. This creates an electric dipole that rotates around the orbit plane at a frequency of $\omega - \omega_{eg}$, after making the rotating wave approximation. The $R_1$ pulse makes the following transformations depending on the state of the entering atoms:

$$|e\rangle \to \frac{1}{\sqrt{2}}(|e\rangle + |g\rangle) \text{ and } |g\rangle \to \frac{1}{\sqrt{2}}(-|e\rangle + |g\rangle) \quad (9)$$

Notice the π phase difference of the superposition states in equation 9, which depends on the initial state of the atom. The choice for which state receives the negative sign is arbitrary. This shift can be visualized on the bloch sphere, which is an alternative way to view the rotating dipole, as shown below the debroglie wavelength representation of fig. 5. The negative signs in equations 7, 8, and 10 can be visualized in a similar way.

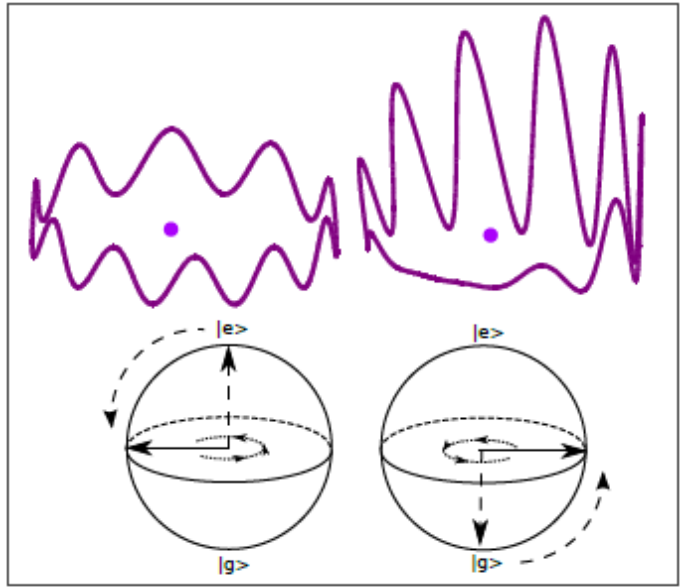

**Figure 5.** DeBroglie wavelength representation of a Rydberg atom with a uniform amplitude and a superposition state (top left and right) with a node and antinode giving rise to an atomic dipole rotating around the plane. Included is the Bloch sphere representation (bottom left and right) for the R1 pulse applied to atoms entering in either e or g states respectively.

After a time delay T, the $R_2$ pulse makes the transformations:

$$|e\rangle \to \frac{1}{\sqrt{2}}(|e\rangle + e^{i\phi}|g\rangle) \text{ and } |g\rangle \to \frac{1}{\sqrt{2}}(-e^{-i\phi}|e\rangle + |g\rangle) \quad (10)$$

Where $\phi=\Delta\omega T$ is the accumulated phase difference between the source S and the atomic coherence during time T. Essentially, there are two different "paths" through the interferometer, and the phase difference between these paths can be scanned in order to detect the e to g transition probability. One should realize there are no two paths. In reality, the atomic state is split in a Hilbert space, with the phases described by a complex exponential. While the Mach-Zender interferometer tests spatial phase relations, the Ramsey interferometer is used for temporal relations. As you can see from figure 4, the transition probability varies sinusoidally, depending on the phase difference. If the atom couples with a cavity mode between the $R_1$ and $R_2$ pulses, the phase and amplitude of the fringes are perturbed. This allows one to probe the details of the atom-cavity coupling process.

## VII. Origins of Phase Shifts

An atom crossing a cavity with a discrete number of photons can be modeled as a classical optical system—light passing through a transparent medium. If the photons in the cavity are detuned from a transition, they will not be absorbed by the atom. However, this non-resonant field in the cavity introduces a phase change in the atomic dipole. Reciprocally, the phase change is experienced by the field as well.

Let me introduce these phase shifts for both cases. While the atoms do not absorb the cavity radiation, they do experience a shift in their atomic energy levels, in effect changing the rotating frequency of the dipole (Haroche, 2013; Haroche, 1998). As the atom exits the mode, the dipole resumes its frequency prior to the interaction. Consequently, the dipole is phase shifted. This deBroglie phase shift can be understood by considering an excited state e atom sent across a cavity with n photons. By making an optical analogy, the phase change results from taking the integral of the potential energy along the atom's path.

The wave function for the system of an e atom with a deBroglie wavelength $\lambda_m$ (being much smaller than the packet width) entering a slightly detuned cavity with n photons can be compared with the Maxwell wave equation for the classical optical system. The potential barrier of the photon field is surmounted if the atomic kinetic energy is large enough. A small portion of the atomic wave packet will be reflected. However, this portion can be ignored for atoms in the beam, which are typically travelling in the hundreds of m/s. The energy equations are:

$$E_{n,k} = \left(n + \frac{1}{2}\right)\hbar\omega + \frac{\hbar\omega_0}{2} + \frac{\hbar^2 K^2}{2M} \quad (11)$$

$$E_{\pm n}(r) = (n+1)\hbar\omega \pm \frac{\hbar}{2}|\delta| \pm \hbar\frac{\Omega^2(r)(n+1)}{|\delta|} \quad (12)$$

where $E_{n,k}$ is the system's total energy, and $E_{\pm n}(r)$ is the perturbation energy as a function of position caused by the nonresonant light shifts, for positive or negative detunings. In equation (11), the first term is the energy of the oscillator, the second term is the e to g transition of the atom, and the last term is the atomic kinetic energy with M being the atomic mass and the wavenumber $K = 2\pi/\lambda_m$. Equation (12) follows from the Jaynes-Cummings model, with $\Omega$ being the Rabi frequency, and $\delta$ the detuning frequency.

Keeping the optical analogy in mind, the light is replaced with a matter wave, and the index follows from the cavity effects. After some manipulation, following the method of Haroche and Raimond (1994), the phase shifts are:

$$\Delta\Phi_e(n) = (n+1)\varepsilon, \quad (13)$$
$$\Delta\Phi_g(n) = -n\varepsilon, \quad (14)$$

With $\varepsilon$ being the phase shift per photon by the cavity on the deBroglie wave, defined as:

$$\varepsilon = \frac{1}{\hbar} t_i \frac{\omega}{\delta} \frac{d^2}{2\epsilon_0 V} \quad (15)$$

with $t_i = l_{cav}/v$, d the matrix element of the dipole operator, and V the mode volume (Haroche and Raimond, 1994).

There is a corresponding phase shift in the cavity field as well. In this case, we treat the atom as an index material. The phase shift caused by a g atom is given by:

$$|\Phi_g\rangle = C_n e^{-in\varepsilon}|n\rangle = |\alpha e^{-i\varepsilon}\rangle \quad (16)$$

Here $C_n$ is a probability amplitude represented as $(\langle n|\alpha\rangle)$, where $|\alpha\rangle$ is a coherent state. Thus, the coherent field undergoes a phase shift of $-\varepsilon$. And for an atom in the e state,

$$|\Phi_e\rangle = C_n e^{i(n+1)\varepsilon}|n\rangle = e^{i\varepsilon}|\alpha e^{i\varepsilon}\rangle \quad (17)$$

With a phase shift of $+\varepsilon$. As you can see, there is an inseparable relation between the field and atom phases, depending on the specific atom-cavity non-resonant interaction. This relation is understood by considering the atom as a wave experiencing dispersive effects from the cavity field, and then seeing the atom as having dispersive effects on the cavity field (Fox, 2006; Haroche and Raimond, 1994).

## VIII. QND measurement of the birth and death of a photon

Most experiments involving the detection of single photons are dependent on their destruction. These detection processes make use of the photoelectric effect, whereby incident photons transmit their energy into an electrical signal. By using a Ramsey set-up, Gleyzes et al. (2007) were able to record the spontaneous birth of a thermal photon. They tracked its lifetime, until its eventual decay from the cavity. In contrast to the photodetection processes, the Ramsey scheme does not disrupt the photon's existence. Again, an atomic beam is sent across the cavity. A π/2 pulse in R1 creates a superposition between the e and g states. The cavity is slightly off resonant with a transition, and is cooled to 0.8K. At this temperature, there is a 5% chance of having a single photon within the cavity at any given time. As discussed, the atom-cavity interaction causes a shift in the phases of the rotating atomic dipole and cavity-field. The shift is dependent on the field intensity, δ, and $t_i$. The last two parameters are adjusted so that the R2 pulse collapses the atom into its e state if a photon is in the cavity. The e state signals the $|1\rangle$ photon Fock state, while a detection of an atom in g signals the $|0\rangle$ vacuum state.

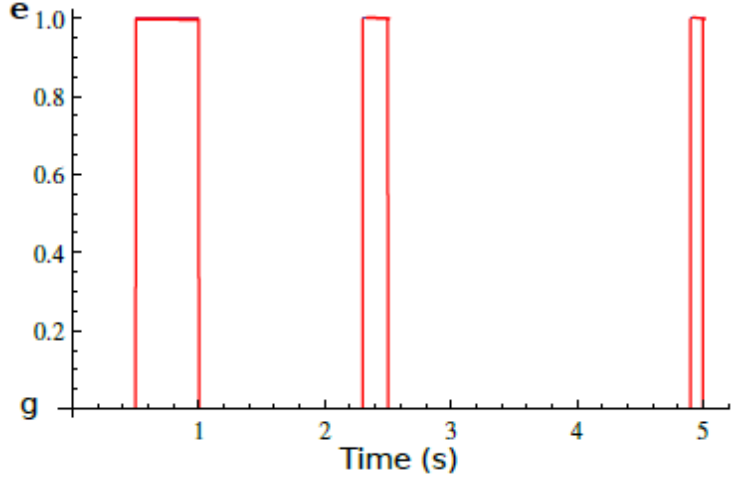

**Figure 6.** Results of a Quantum Non-Demolition experiment of a single thermal photon in a slightly detuned cavity. Detection of an e atom signals the presence of a photon. See (Gleyzes et al., 2007) for experimental data.

As you can see from fig. 6, the first photon remains in the cavity for an unusually long time, almost 0.5 sec. We can see two other photon events. It appears that the fluctuations diminish with time. This is exactly what we would expect for a thermal field, considering "Maxwell's Demon." This is an unprecedented QND measurement of a single photon. Refer to Nogues et al. (1999) for a similar experiment.

## IX. Quantum Phase Gate

The method described above for detection of a single photon can be compared with that of the quantum phase gate. The latter is an important component for quantum computation. We saw in sec. V that a 2π Rabi pulse induced on an atom within the cavity produces a phase shift of π, and have considered e and g Rydberg states corresponding to n=51 and 50 respectively. With the single e,g resonant cavity mode in mind, let us now introduce a third level, called i for n=49. This i state is far from resonance and does not interact with the cavity mode. Single atoms are sent through the cavity with a velocity selected for a 2π interaction. If the atom enters in state i or the cavity is empty, no phase change occurs. The π-shift occurs for an atom in g with one photon in the cavity. For an atom in a superposition of g and i, the system is unchanged if the cavity is empty, and shifted by ϕ=π when the cavity contains one photon given by:

$$(c_g|g\rangle + c_i|i\rangle)|1\rangle \rightarrow (e^{i\phi}c_g|g\rangle + c_i|i\rangle)|1\rangle \qquad (18)$$

In equation 18, $c_g = c_i = 1/\sqrt{2}$. . If the field is in a superposition of zero and one photon states, the phase is unchanged by an atom in i, while for g we have:

$$|g\rangle(c_0|0\rangle + c_1|1\rangle) \rightarrow |g\rangle(c_0|0\rangle + e^{i\phi}c_1|1\rangle) \qquad (19)$$

Again, $c_0 = c_1 = 1/\sqrt{2}$. One can see that if both field and atom qubits are in state superpositions, the gate creates atom-cavity entanglement. To test equation 18, the cavity is prepared with either zero or one photons. For the second case an atom is prepared in the e state, and sent through the cavity, where it undergoes a π-interaction in order to add one photon to the cavity. A second atom is then sent, first being prepared in a g and i state superposition by a π/2 ramsey pulse. It then undergoes a 2π rotation through the cavity. A second π/2 ramsey pulse is applied, before detection by field ionization. The Ramsey fringes for this experiment are shown in fig. 7(a). The π shift is clearly apparent.

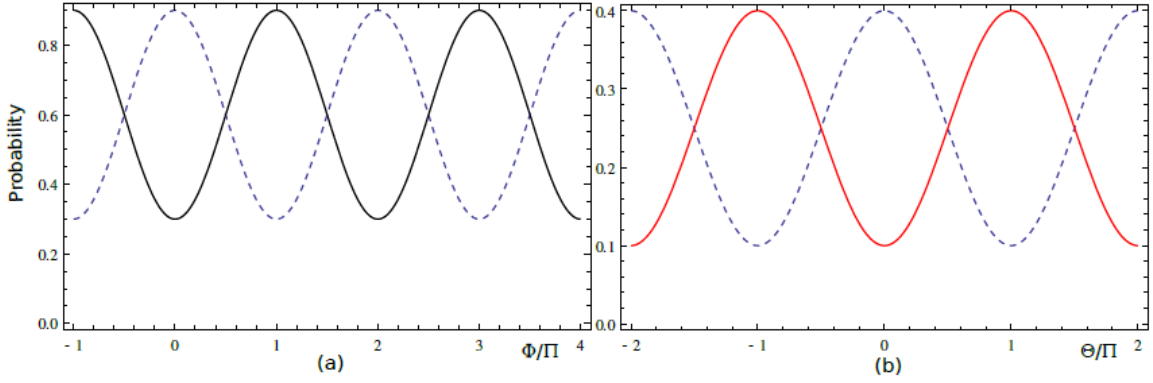

**Figure 7.** In (a), we have the probability of detecting the second atom in state g versus the phase of the Ramsey interferometer. The solid line corresponds to the cavity in vacuum and the blue dotted line for one photon. (b) shows the conditional probabilities of detecting the third atom in e if the second atom was in the g state (red line) or the i state (blue dotted) versus θ. Curves are generated to match the data from (Raimond, Brune, and Haroche, 2001).

The test of equation 19 involves the creation of a coherent field in the cavity, suspended in an equal superposition of the zero and one photon states. The atom is prepared in the same fashion as for the previous test, undergoing a 2π rotation through the cavity. If the atom is detected in i, the field phase should be unchanged. If detected in the g state, the field in the cavity evolves according to:

$$c_0|0\rangle + c_1 e^{i\phi}|1\rangle \cong |e^{i\phi}\alpha\rangle = |-\alpha\rangle \qquad (20)$$

and the phase is π-shifted. This shift is detected by injecting into the cavity a second coherent source with amplitude $\alpha e^{i\theta}$, which adds to the field already in the cavity. The phase depends on the detuning between the two coherent fields and the resulting amplitude varies between 0 and 2α as a function of θ. A third atom is then sent initially in g, where it undergoes a π rotation through the cavity with 1 photon. Thus, the probability of finding the atom in e is equal to the probability of finding 1 photon in the cavity. The probabilities for detecting the third atom in e for the cases that the second atom was in i or g with respect to θ, are shown in fig. 7(b). The shift in these signals results from the interference of the two field pulses (Raimond, Brune, Haroche, 2001; Rauschenbeutel et al., 1999). This is another QND method for detecting single photons.

The quantum phase gate is a realization of the two qubit controlled-NOT (C-NOT) gate, seen in table 1. The control qubit is unaltered by the gate, but flips the target qubit conditioned on the state of the control.

**Table 1.** Truth values for the controlled-NOT transform.

| Input qubits | | Output qubits | |
|---|---|---|---|
| Control | Target | Control | Target |
| $\|0\rangle$ | $\|0\rangle$ | $\|0\rangle$ | $\|0\rangle$ |
| $\|0\rangle$ | $\|1\rangle$ | $\|0\rangle$ | $\|1\rangle$ |
| $\|1\rangle$ | $\|0\rangle$ | $\|1\rangle$ | $\|1\rangle$ |
| $\|1\rangle$ | $\|1\rangle$ | $\|1\rangle$ | $\|0\rangle$ |

The presence of a photon in the cavity is the control qubit in the state $|1\rangle$ and the passing atoms are the targets. The target is flipped if and only if the control is $|1\rangle$. In theory, a quantum computer could be constructed solely from the operations described in sec. V and the quantum phase gate (Protsenko, Reymond, Schlosser, & Grangier).

## X. Discussion

Those well-versed in the field will see this is in no way an exhaustive text. The goal was to convey the basic physics and just a few striking experiments that have had a significant impact. Many other related experiments involving three particle entanglement, non-resonant entanglement, and detection of decoherence have not been included in this work. For more information on these studies, refer to references (Raimond, Brune, Haroche, 2001; Pellizzari et al., 1995; Brune et al., 1996). The main ingredients for Cavity QED have been presented, and would suffice for further inquiry.

The strong coupling regime has offered strong hope towards greater control over quantum systems and their implementation in computation and cryptography. However, the transient lifetime of prepared entangled states is a limiting factor towards progress. As the computational schemes increase in number of operations, the need to lengthen the coherence time of the system is of primary importance. Many are trying to further enhance the atom-cavity coupling and reduce photon decay. If one was to measure decayed photons, unintended consequences would destroy the prepared coherence (Haroche, 1998). This is because the decayed photons are still entangled with the atomic qubits, even though they have escaped the closed cavity. In an attempt to lengthen the coherence time, efforts to achieve quantum feedback mechanisms are currently being pursued (Mabuchi & Doherty, 2002). These feedback loops involve measuring the photons within the cavity using QND methods, and then coherently adding photons in order to preserve the Fock states (Peaudecerf et. al, 2013). A practical application of QND methods could be the development of new single photon detectors. The current technologies rely on the destruction of the detected photons, causing a loss of potentially useful information. In the QND measurements, the photon itself is preserved.

The issue of escaped photons from the cavity is more a technological problem, and could be considered as separate from the fundamental consequences of decoherence. When a state superposition decoheres, all of the quantum weirdness is lost to the environment. The states' evolution in its prepared basis becomes entangled with the numerous environmental degrees of freedom. Furthermore, if one superposition state is entangled with another, the interaction with the environment breaks the entanglement. The states are no longer non-locally related. The damping seen in figure 3 illustrates a signature of decoherence, which is the transfer of a quantum state into a statistical mixture. However, in that case, the environment acts on the measurement apparatus, and not on the quantum object. The disparity between quantum theory and observation of its effects, at least in the cavity QED context, is an issue over the inseparable system and environment coupling. A modification of the theory sometimes comes to mind when trying to address the effect of the observer on the state of the system. Could there be a meaningful relationship between the effect of the observer (i.e. the measurement apparatus) and the decohering effect of the immediate environment? It appears that most interpretations of quantum mechanics disregard any distinction between the two. If the space (including the particles) of the environment have the same quantum mechanical properties as those of the measurement apparatus, as they should, then what is it about the observer that restricts the system to a particular value? The environment has a similar restrictive effect. But instead of acting instantaneously, its effects range in time. This suggests that there may be an unaccounted factor which deterministically acts on the quantum system beyond what is presented by decoherence theory. Many are reluctant to invoke consciousness into the measurement problem, and this is understandable. But this does not mean that the measurement apparatus, and the means by which the measurement is carried out, are not involved with how a superposition can jump into a well-defined state. The experimental conditions offered by Cavity QED schemes will hopefully be able to elucidate these questions and other controversial issues over the fundamental laws of physics.

## ACKNOWLEDGEMENTS

I would like to thank Chad Orzel for giving many useful comments on a draft of this work, Gary Reich for a discussion on Maxwell's equations, and Steve Ulin for helping with constructing Figures 2 and 5.